\documentstyle[twocolumn,prl,aps]{revtex}

\begin{document}

\twocolumn[\hsize\textwidth\columnwidth\hsize\csname
@twocolumnfalse\endcsname

\title{Two-Stage Rotational Disordering of a Molecular Crystal Surface:
C$_{60}$}

\author{C. Laforge$^{1,4}$, D. Passerone$^{2,4}$, A.B. Harris$^{3}$,
   Ph. Lambin$^{1}$ and E. Tosatti$^{4,5}$}
\address{1.Facult\'es Universitaires Notre-Dame de la Paix, Rue de Bruxelles
  61, B-5000 Namur, Belgium.\\
2.Max Planck Institut fuer Festkoerperforschung, Heisenbergstr. 1,
  D-70569 Stuttgart, Germany. \\
3.Department of Physics, University of Pennsylvania,
  Philadelphia, Pennsylvania 19104-6369\\
4.International School for Advanced Studies (SISSA), and Istituto
 Nazionale Fisica della Materia, via Beirut 2-4, I-34014 Trieste, Italy \\
5.International Centre for theoretical Physics, (ICTP), P.O. Box 586,
I-34014 Trieste, Italy}

\date{today}

\maketitle

\begin{abstract}
We propose a two-stage mechanism for the rotational surface disordering 
phase transition of a molecular crystal, as realized in C$_{60}$ fullerite.  
Our study, based on Monte Carlo simulations, uncovers the existence 
of a new intermediate regime, between a low temperature ordered 
$(2 \times 2)$ state, and a high temperature $(1 \times 1)$ disordered 
phase. In the intermediate regime there is partial disorder, 
strongest for a subset of particularly frustrated surface molecules.
These concepts and calculations provide a coherent understanding of 
experimental observations, with possible extension to other molecular 
crystal surfaces.

\end{abstract}

]

Surface-specific thermal disordering phenomena such as for instance surface
melting \cite{surfacemelting}, are of considerable interest. Conceptually, 
they demonstrate how a chunk of the metastable phase (e.g, the liquid 
phase within the solid domain of stability) can be imprisoned in full 
equilibrium by restricting it to an interface region of space. Practically, 
they provide at a temperature below the triple point a germ of the disordered 
phase that makes a major difference to all surface-related properties, as 
well as to the bulk transformation kinetics. 

In molecular crystals, thermal surface disordering is richer than in atomic 
solids. Because of the presence of rotational degrees of freedom, the bulk 
develops rotational and positional melting transitions that are 
separate, and both of which might possess a surface counterpart. One is thus 
led to seek the existence of a well defined surface rotational melting for a
molecular crystal. However until recently, very little phenomenology has
gathered addressing that issue. 

That situation recently changed after the discovery of the C$_{60}$ 
molecule\cite{kroto} and its subsequent availability in macroscopic 
quantities\cite{CD-2} in its crystalline form, the fullerite. Here 
the near sphericity of the molecules favors rotational disordering at 
a relatively low temperature, when positional disorder and evaporation 
are still irrelevant owing to a sufficiently large cohesion~\cite{CC-9}. 
On the theoretical side, the remarkable symmetry of the molecule leads to 
considerable simplifications with the help of symmetry adapted functions
\cite{CC-24}, such as those originally used for the CD$_{4}$ molecules
\cite{JEK}.

Bulk measurements showed that  C$_{60}$-fullerite undergoes
a first-order rotational melting transition at 260K, characterized by fixed
orientations of the molecules in the space group $Pa\bar{3}$ at
low temperature transforming to a plastic phase with rotational
isotropy, space group  $Fm\bar{3}m$, at high temperature
~\cite{CC-a,CC-b,CC-12}. In the $Pa\bar{3}$ phase, the fullerite
structure is composed of four sublattices. Neutron powder diffraction
studies~\cite{CC-59} of this phase have revealed the orientational
configuration of the C$_{60}$ molecules. The optimal orientation
is such that for all adjacent couples of C$_{60}$, an electron-rich
double bond of one molecule faces an electron-poor pentagon of the other. 
In the $Fm\bar{3}m$ phase, the molecules are still not completely 
free to rotate as they are still correlated, and moreover they feel 
a crystal field of cubic symmetry. However, in this phase angular 
correlations are short-ranged, and molecules are all equivalent, 
as opposed to the $Pa\bar{3}$ phase, with four inequivalent molecules 
and long-range angular correlations.

The structure of a free (111) surface of fullerite can differ from
that of a bulk (111) plane due to the lowering of the number
of neighbors.  The first experiments showing a specific surface
disordering were high-resolution reflection electron-energy-loss
spectra (EELS) obtained by Goldoni et al~\cite{CC-27}. This technique
is sensitive to the surface layer only due to the short escape
depth of the scattered electrons. The authors reported an abrupt
jump around 230K in the energy of the C$_{60}$ triplet exciton.
This discontinuity is an indication that the fullerite (111) surface
undergoes the rotational disordering phase transition at a temperature
some 30K lower than the bulk. A similar transition temperature
($230\pm 20$K) has been observed in LEED studies of C$_{60}$ films 
\cite{Ben} but the incertitude in the temperature was too large to 
interpret it as a specific surface effect.

At first sight, such a lowered surface transition temperature is reminiscent
of ordinary, positional surface melting of solids, where surface premelting
precedes and initiates the bulk transition~\cite{surfacemelting}.
However, it soon appeared that in fullerite the surface disordering
took place in the topmost layer only~\cite{CC-27,CA-24}, in contrast
with standard surface melting, where the liquid film grows progressively
invading the crystal. At the same time, a conceptual difference was also
noted. Positional order in a crystal is forced exclusively by interparticle
interaction; rotational ordering instead involves both intermolecular
angular interactions and a cubic crystal field that would be present
and effective on each molecule even if all the surrounding molecules were
perfect spheres. A surface molecule thus feels not only less rotational
interactions, but also a crystal field which may generally frustrate 
bulk order\cite{CC-26}. On the other hand, due to the weak short range 
nature of intermolecular potentials, the molecules in the second layer, 
immediately below the surface are already very close to a bulk situation. 
The system, between 230K and 260K, can thus be regarded as an ordered 
solid with a disordered surface.

We now underline the fact that experimental data suggest in fact not one,
but {\em two} surface transitions, the second one taking place at even lower
temperature. There is a less-marked discontinuity around 160K in the triplet
exciton energy revealed by EELS~\cite{CC-27}. It was explained as a
consequence of the molecular rotation relaxation time which at this
temperature becomes similar to the exciton lifetime.  However an 
independent evidence for a transition or crossover at that lower 
temperature is the presence around 160K of a shoulder in the LEED 
intensity of the $Pa\bar{3}$ diffraction spots versus temperature, 
absent in the bulk X-ray intensity~\cite{CC-29}.  Helium atom
scattering (HAS) diffraction studies of the (111) surface of fullerite by
Glebov et al\cite{CC-57} clearly showed a change from a (2X2) to a (1X1)
structure taking place at 235K.  But, in addition, the behavior of the
Debye-Waller Factor (DWF) of the (0,0) Bragg spot between 140K and 235K was
completely unexpected. Instead of decreasing monotonically with increasing
temperature, the DWF started increasing around 140K and reached a maximum
at 235K. This was explained as a consequence of preempted disordering of
C$_{60}$ molecules at low coordination defect sites.  This behavior, 
found to be reversible, cannot be explained if the surface layer were 
totally ordered below 235K.

In the present work, we studied the surface orientational ordering of
fullerite by classical Monte-Carlo (MC) simulation. We considered
a C$_{60}$ molecule as a rigid body, and used the potential developed
by Lamoen \cite{CC-4} to model the intermolecular interactions. That
model considers 210 interacting centers on each fullerene: 60 are
localized on the carbon atoms, 60 are at the center of the single
bonds, and three are distributed along each of the 30 double bonds
to take their spatial extension into account. The repulsive part of
the intermolecular potential is a sum of Born-Mayer terms between all 
the interacting centers (c) and the attractive part is made of $R^{-6}$ 
van der Waals interactions between the atoms (a)

\begin{equation}
V=\sum_{a_{1}a_{2}}\frac{A}{r^{6}_{a_{1}a_{2}}}
+\sum_{c_{1}c_{2}}B_{c_{1}c_{2}}
\exp \left[-C_{c_{1}c_{2}}r_{c_{1}c_{2}}\right]
\end{equation}

Our MC simulations were based on a standard Metropolis, acceptance-rejection
algorithm. The molecules were placed in a semi-infinite fcc lattice with a
(111) surface. We considered twelve non-equivalent molecules, four in the
first layer (the surface layer), four in the second layer, and four in the
third layer. All other molecules in the system were assumed to be copies of
this twelve-molecules basic cell, with periodic boundary conditions applied in
two dimensions. In the direction of the bulk, the second layer was reproduced
at the fourth layer, shifted horizontally in order to simulate the fcc
stacking. The center of mass of the molecules were held fixed to lattice
sites, while the temperature dependent lattice parameter of the structure was
set to the experimental value\cite{CC-7}.  The twelve molecular orientations
were allowed to change randomly.  We performed a complete calculation of the
energy at each MC step.

Earlier theoretical studies of the rotational bulk phase transition have shown
the usefulness of a set of order parameters based on the cubic harmonics
\cite{CC-24}. We used an efficient and original algorithm to generate these
cubic harmonics. Starting from generating functions of the irreducible
representations $k$ ($A_1$, $A_2$, $E$, $T_1$, $T_2$) of the cubic group
$O_h$, it is possible to expand them into Taylor series and to extract from
that expansion the number of irreducible representations for each manifold $l$
of the intermolecular potential. Using the generators $\Gamma\left(k\right)$
of each irreducible representation, one can construct automatically the cubic
harmonics\cite{patera}
\begin{equation}
\left[\Gamma^{\left(k\right)}\left(g\right)\right]
E^{\left(l,k\right)}\left(\vec{r}\right)=
E^{\left(l,k\right)}\left(\Gamma^{\left(T_{1}\right)}\left(g\right)
\vec{r}\right)
\end{equation}
where $g$ belong to $O_{h}$ and $E^{\left(k,l\right)}$ is the integrity 
basis of order $l$ for the irreducible representation $k$. Since
the cubic harmonics must be homogeneous polynomials of the coordinates, 
this system can be solved easily, and one obtains analytical expressions
for all the cubic harmonics, to be used with the atomic coordinate of 
the 60 atoms of the molecule. In particular, the following expression 
was calculated
\begin{equation}
h_{n}^{\left(l,k\right)}=
\sum_{a=1}^{60}E^{\left(l,k\right)}\left(\vec{r_{a}^{n}}\right)
\end{equation}
where $\vec{r_{a}^{n}}$ is the position vector of atom $a$ in the $n^{th}$
molecule in the unit cell. Using the C$_{60}$ atomic coordinates automatically
introduces the icosahedral symmetry of the molecules.  Another advantage of
this method is to naturally allow for a future inclusion of internal molecular
vibrations, not treated at this stage. Due to the icosahedral symmetry,
$h_{n}^{\left(l,k\right)}$ differ from zero for $l$ = 6, 10, 12, 16, 18
$\cdots$ . We used $l=10$ throughout. For each molecule (index $n$), we
calculated a scalar order parameter
\begin{equation}
\eta_{n}^{\left(l,k\right)}=|\left<h_{n}^{\left(l,k\right)}\right>|^{2}
\end{equation}
where $\left<h_{n}^{\left(l,k\right)}\right>$ is an average over
the configurations generated by the MC algorithm. For all
the irreducible representations of the $O_{h}$ group, this parameter
was traced as a function of the temperature to observe the phase
transition as it has the largest amplitude in the ordered phase.

As mentioned in~\cite{rapc} $T_{1g}$ and $T_{2g}$ are the two primary order
parameters.  We adopted the irreducible representation $T_{2g}$ as the best
indicator of the rotational phase transition, as the corresponding order
parameter vanishes in the disordered phase~\cite{CC-24}. Figure 1 shows the
temperature dependence of that order parameter for the twelve molecules of our
system. The eight bulk-like molecules of layers 2 and 3 (represented by
triangles), plus all their infinite copies, undergo a clear phase transition
at 250K. This value is affected by size, in particular by artificial
correlations built in a small system like the one we considered. By
comparison, the mean-field theory transition temperature was however even
smaller, namely 200K~\cite{CC-4}. However, only the dominant part of the
multiple order parameters was originally used in the mean-field molecular
equations of ref.\,\onlinecite{CC-4}. An extension of the number of order
parameters did raise the mean field transition temperature to
230K~\cite{CC-12}.

At low temperature, the molecules in the (111) surface (represented
by circles in figure 1) keep the same orientation as in the bulk, 
corresponding to the $(2 \times 2)$ superstructure observed in HAS
diffraction. In this configuration, shown in figure 2, one can immediately 
distinguish two kinds of molecular orientations. Three molecules in
the surface $(2 \times 2)$ unit cell have only a two-fold axis normal to the
surface and the $Pa\bar{3}$ threefold axis 30 degrees off. The fourth
molecule, which we denote as BS, has instead the three-fold axis normal to 
the surface. Calculations performed by rotating one molecule while 
keeping a fixed orientation of all the other molecules show that 
the ``crystal field'' angle-dependent potential felt by the rotating 
molecule is different, and shallower for the BS molecule than for the 
three others. This suggests that the rotational thermal behavior of the BS 
molecule must differ from that of the other surface molecules~\cite{CC-26}.

In fact with increasing temperature, the four surface molecules do behave
differently. The BS practically becomes a quasi-free rotator above around 
150K whereas the other three molecules do so only above a much higher phase
transition temperature, which is around 230K.

In the low temperature phase, the surface symmetry is threefold due to the
local threefold axis of the BS molecule. At 150K, we observe a dramatic
reduction of threefold order with strong rotational disordering of the BS
molecule. Although strictly speaking the threefold order of that molecule
cannot vanish, it does however become so small to assimilate it to a free
rotator, in which case the surface symmetry would become sixfold because the
BS molecule environment is characterized by a sixfold axis.  Thus 150K
represents a crossover temperature, where there is no strict change of surface
symmetry and probably no proper surface phase transition, but there is
nonetheless a change from essential order to essential disorder for BS
molecules only.

Hence our system exhibits altogether four phases or regimes  (see table I):

(a) a fully ordered $Pa\bar{3}$ state (effective $p3$ surface symmetry) 
below 150K.
(b) an intermediate regime between 150K and 230K with one essentially
disordered and three ordered surface molecules per unit cell ($p6$
effective symmetry of the surface layer) over a $Pa\bar{3}$ ordered bulk.
(c) a disordered surface layer on an ordered $Pa\bar{3}$ bulk between 230K and 250K.
(d) a fully rotationally disordered $Fm\bar{3}m$ ($p6m$ symmetry of the 
surface layer) above 250K.

In the simulations, full thermal disordering of the molecules takes place at
the topmost layer only in agreement with evidence from EELS~\cite{CC-27}.  In
addition, the main surface transition temperature (230K) agrees very well with
the observations by LEED, EELS and HAS.  The intermediate BS-disordered regime
between 150K and 230K can explain the weaker experimental anomalies mentioned
above, in particular the feature at 160K in the triplet exciton energy,
signaling in our interpretation an approximate change of surface symmetry from
$p3$ to $p6$.

We note that in LEED and HAS diffraction, this change is not picked up since
the unit cell remains $(2 \times 2)$ up to 230K.  We suggest instead that STM
experiments be carried out in the appropriate temperature interval to test in
more detail our predicted two-stage surface disordering. Our model provides a
plausible explanation for the behavior of the (0,0) Debye-Waller factor in HAS
diffraction~\cite{CC-57}, its increase between 140 and 235K due to the BS
molecule rotating freely in that temperature interval. Due to the resulting
partial disordering of the $(2 \times 2)$ surface cell, the intensity of some
diffracted beams is reduced and transferred to the specular
spot~\cite{heliumsc}, which compensates the thermal attenuation of the elastic
reflection~\cite{CC-57}. Our proposed selective thermal disordering of BS
molecules may also provide an explanation of the anomalous features observed
near 160K in X-ray powder diffraction intensities\cite{M-1,M-2}.  Such
anomalies were also observed by David~\cite{D-1} in high-resolution neutron
scattering. If these anomalies were not reported later on, it is probably due
to the increasing quality (and then the grain size) of the C$_{60}$ powder,
decreasing the surface/bulk ratio.

In summary, the study of fullerite may prove of paradigmatic importance
to a much broader class of molecular crystals. Our study indicates that 
rotational thermal surface melting of a molecular crystal differs from 
ordinary, positional surface melting, in one important aspect, with 
several consequences. The rotational ordering forces comprise both crystal 
fields and intermolecular interactions, while the former is missing in the 
positional case. At the surface, both interactions and crystal fields are 
different from the bulk. When, as in fullerite, the crystal field is 
quantitatively very important in the bulk, its heavy change at the surface 
may lead to an early thermal disordering taking place strictly in the first
layer, as in all other layers the crystal field is bulk-like. Moreover,
crystal fields may differ so much between one surface molecule and another, 
to cause surface thermal disordering to take place in separate stages,
the weaker molecules disordering much earlier than the stronger ones.

The authors acknowledge support through the EU project FULPROP, contract
number ERBFMRXCT 970155. This research was also funded by NSF and the PAI
P4/10 on reduced dimensionality systems. Work at SISSA was sponsored by MURST
COFIN99, and by INFM/F . The authors thanks also P. Senet, V. Meunier, L
Henrard, P. Rudolf, S. Modesti, and A. Goldoni for many interesting
discussions.

\begin{table}
\begin{center}
\caption{Ordering phase transitions in C$_{60}$-fullerite with a (111)
  surface.}
\begin{tabular}{clll}
temperature & bulk & surface & surf. sym.\\
\hline
below 150K & ordered & ordered & p3 \\
150K-230K & ordered &
$\left\{ \begin{tabular}{l}
{\rm \small 1 of 4 mol. disordered} \\
{\rm \small 3 of 4 mol. ordered}
\end{tabular} \right.$ & p6\\
230K-250K & ordered & disordered & p6m \\
above 250K & disordered & disordered & p6m
  \\
\end{tabular}
\end{center}
\end{table}

\begin{figure}
\caption{Variation of the order parameter $\eta_{n}^{\left(l,k\right)}$
corresponding to $l=10$ and $k=T_{2g}$ versus temperature. The
curves correspond to the three behaviors observed in our Monte-Carlo
simulations. The triangles are the arithmetic means of the order parameters
associated to the bulk molecules. The circles are the arithmetic means of all
the surface molecules but the molecule BS which has a behavior pointed out by
the squares.}
\end{figure}

\begin{figure}
\caption{Equilibrium configurations of the C$_{60}$ fullerite in the (111)
plane with space group $Pa\bar{3}$. The unit cell is $(2 \times 2)$. Three of
the four molecules have a two-fold axis normal to the plane. The fourth has a
three-fold axis. This last molecule experiences a six-fold symmetry
interaction from its neighbors located in the same layer.}
\end{figure}

\end{document}